\documentclass [12pt]{iopart}
\usepackage {iopams}
\usepackage{setstack}
\usepackage{amstext}
\usepackage{dsfont}
\usepackage{amsfonts}
\usepackage{graphicx}
\usepackage{subfigure}
\usepackage{array}
\usepackage{appendix}
\usepackage{mathrsfs}
\usepackage{color}

\usepackage{adjustbox,lipsum}
\usepackage{upgreek}
\usepackage{algorithm}
\usepackage{algpseudocode}
\usepackage{makecell}
\usepackage{adjustbox,lipsum}
\usepackage{bm}

\usepackage{amsthm}
\usepackage{amssymb}
\usepackage{latexsym}
\usepackage{hyperref}
\usepackage{float}
\usepackage{verbatim}
\usepackage{bbold}
\usepackage[normalem]{ulem}

\usepackage{upgreek}

\usepackage{makecell}
\usepackage{adjustbox,lipsum}

\usepackage{algorithm}
\usepackage{algpseudocode}

\newcommand{\ket}[1]{\left|#1\right>}
\newcommand{\abs}[1]{\left|#1\right|}

\newtheorem{theorem}{Theorem}
\newtheorem{proposition}{Proposition}
\newtheorem{lemma}{Lemma}
\newtheorem{corollary}{Corollary}
\newtheorem{definition}{Definition}

\begin{document}
\title[]{Secure PAC Learning: Sample‑Budget Laws and Quantum Data-Path Admissibility}

\author{Jeongho~Bang$^{1,2}$}

\address{$^1$ Institute for Convergence Research and Education in Advanced Technology, Yonsei University, Seoul 03722, Republic of Korea}
\address{$^2$ Department of Quantum Information, Yonsei University, Incheon 21983, Republic of Korea}

\vspace{10pt}

\begin{indented}
\item {\em Correspondence and requests for materials should be addressed to J.B}
\end{indented}

\ead{\mailto{jbang@yonsei.ac.kr}}

\begin{abstract}
Security in machine learning is fragile when data are exfiltrated or perturbed, yet existing frameworks rarely connect data-path security to learnability. In this work, we develop an operational theory of secure learning grounded in the probably-approximately-correct (PAC) viewpoint. An explicit stopping time combines the event that a trained hypothesis attains a target accuracy with the event that a run-based validation gate halts within a finite sample budget. We derive a closed-form sufficient budget requirement for this joint PAC-within-budget guarantee under an admissible random-classification-noise channel. We then specialize the construction to a BB84-like quantum label path. Under the stated ideal single-qubit, authenticated-classical-channel, memoryless, basis-symmetric, collective-attack, asymptotic, and one-way-reconciliation assumptions, the standard Holevo bound gives the protocol-specific information-advantage criterion $1-2h(\eta)>0$, with the threshold $\eta_{\mathrm{BB84}}\simeq0.11$. The quantum layer is not invoked to reduce distribution-free PAC sample complexity; rather, it turns a designer-chosen classical noise tolerance into a physically testable, protocol-dependent security condition by linking information acquisition to observable disturbance. Below the threshold, the PAC and information-advantage conditions provide complementary statistical and physical guarantees; at or above it, the present protocol and proof no longer certify the latter. Basis sifting is incorporated explicitly in the conversion from sifted-sample budgets to expected raw channel uses.
\end{abstract}

\maketitle

\section{Introduction}\label{Sec:1}

Security is no longer an afterthought in machine learning~\cite{Barreno2010}. Deployed systems routinely face threats that arise before the learning begins---during data acquisition and transport---where exfiltration of raw streams and subtle tampering of labels and/or features can undermine reliability. Concrete examples abound: targeted data poisoning and backdoor insertions that push a learner toward attacker-chosen behaviors while preserving high apparent accuracy on held-out data~\cite{Biggio2012,Biggio2018,Jagielski2018}; privacy-oriented attacks such as model inversion and membership inference that expose sensitive attributes or the presence of individuals in training sets~\cite{Fredrikson2015,Shokri2017}; data extraction that reconstructs learned parameters or even verbatim training samples~\cite{Carlini2021}. While some cryptographic primitives (e.g., authenticated channels, secure aggregation) and the worst-case adversarial robustness provide essential building blocks, they rarely furnish a methodology for defining and analyzing the security on the same footing as the learnability. This motivates a theory of secure machine learning that binds channel properties, information leakage, and statistical guarantees into a complete learning framework.

At its core, learning theory asks when a learner trained on samples will generalize to the population. The probably-approximately-correct (PAC) learning has, for decades, provided a flexible language for analyzing accuracy, confidence, capacity, and sample size across diverse algorithms in computational learning theory~\cite{Valiant1984, Langley1996}. Modern refinements extend this PAC learning to high-capacity models, including deep neural networks, through margin-based and compression-based analyses that explicitly link geometry, complexity control, and finite sample budget~\cite{Arora2018}. In particular, the PAC-Bayes program supplies distribution-dependent yet explicit certificates whose posteriors can be tailored to the learned representation~\cite{Seeger2002,Mcallester2003,Dziugaite2017}. These developments complement the uniform convergence and stability analyses~\cite{Bartlett2002,Bousquet2002}, positioning the PAC paradigm as both an asymptotic lens and an experiment-facing framework. The independent random-classification-noise model used below should be distinguished from PAC learning under malicious or adversarial errors, for which the corruptions may be chosen strategically~\cite{KearnsLi1993}.

Quantum extensions sharpen this agenda---which can give rise to quantum-secure machine learning. QML has initially been explored for potential algorithmic speedups or resource savings enabled by quantum properties~\cite{Yoo2014,Rebentrost2014,Biamonte2017}. A distinct direction uses QML as a tool for security tasks, such as quantum anomaly detection~\cite{Liu2018}, while secure quantum pattern communication protects communicated patterns~\cite{Harney2022}. Our focus is instead the security of the learning process and its data path~\cite{Bang2015,Sheng2017,Song2021a,Bang2025}. Encoding classical labels into nonorthogonal quantum states alters the physics of eavesdropping: the no-cloning theorem~\cite{Wootters1982,Scarani2005}, or equivalently the information-disturbance tradeoff~\cite{Fuchs1996}, implies that information acquisition by an eavesdropper is accompanied by disturbance under the stated device and channel assumptions. This disturbance is reflected in the QBER and can bound the eavesdropper's accessible information. Importantly, quantum and classical examples have the same distribution-free PAC sample complexity up to constant factors~\cite{ArunachalamDeWolf2018}; the role of the quantum layer here is therefore security-oriented rather than a generic reduction of the learning sample complexity.

The present work shares with Refs.~\cite{Song2021a,Bang2025} the broad philosophy of using quantum information--disturbance relations to protect learning, but its connection to these studies is conceptual rather than formal. Ref.~\cite{Song2021a} derives a no-broadcasting-based secure-PAC sample condition, while Ref.~\cite{Bang2025} compares the learning quality available to authorized and eavesdropping learners. By contrast, we ask whether a learner--data-path pair can be certified as a secure PAC learning system under a prescribed finite budget. To answer this question, we establish a self-contained framework with distinct definitions, acceptance criteria, and mathematical guarantees, together with a BB84-based physical security certificate. These elements do not follow from refining the earlier formulations. Thus, our present work constitutes an independent extension of PAC theory for secure learning.

\section{Lower bound on the learning confidence in PAC framework}\label{Sec:PAC}

The probably-approximately-correct (PAC) paradigm provides a distribution-free calculus for quantifying when a learner can, with high confidence, return a predictor whose risk is within a prescribed tolerance. Let $\Theta=\{ (x_j, y_j) \}_{j=1}^m$ be an i.i.d. sample drawn from an unknown distribution $\mathscr{D}$ over inputs, write $m := \abs{\Theta}$, let $c \in \mathcal{C}$ be the target concept, and let $\mathcal{H}$ be a hypothesis class. Under the loss, we define the population and empirical risks by
\begin{eqnarray}
R(h,c) := \Pr_{x \sim \mathscr{D}} \big[ h(x) \neq c(x) \big], \quad \widehat{R}_{\Theta}(h) := \frac{1}{m} \sum_{j=1}^{m} \mathds{I} \left[ h(x_j) \neq y_j \right],
\label{eq:risk-defs}
\end{eqnarray}
where $\mathds{I}(\omega)$ is the indicator of an event $\omega$. Given inaccuracy $\varepsilon \in (0,\frac{1}{2})$ and failure probability $\delta \in (0,1)$, a hypothesis $h$ is called \emph{$\varepsilon$-approximate} if $R(h,c) \le \varepsilon$. Here, we adopt the standard notion of the PAC learnability~\cite{Valiant1984,Langley1996}:
\begin{definition}[PAC learnability]\label{def:PAC}
A concept class $\mathcal{C}$ is \emph{PAC-learnable} by a learner (or learning algorithm) $L$ if there exists a polynomial function $\mathrm{poly}(\cdot,\cdot,\cdot)$ such that, for all $\varepsilon \in (0,\frac{1}{2})$ and $\delta \in (0,1)$, every distribution $\mathscr{D}$ over the inputs, every target concept $c\in\mathcal{C}$, and any sample size $m \ge \mathrm{poly}\bigl( \frac{1}{\varepsilon},\frac{1}{\delta},\mathrm{comp}(\mathcal{C}) \bigr)$, $L$ returns an $\varepsilon$-approximate hypothesis $h \in \mathcal{H}$ with probability at least $1-\delta$. Here, $\mathrm{comp}(\mathcal{C})$ denotes a capacity parameter of $\mathcal{C}$ (e.g., $\ln\abs{\mathcal{C}}$ for finite classes or the VC dimension for infinite classes).
\end{definition}

In the noiseless setting---i.e., $y_j=c(x_j)$ and $\mathcal{H}$ contains a perfect classifier---the classical finite-class analysis yields the familiar upper bound on the sample complexity. The argument proceeds by controlling the deviation of $\widehat{R}_{\Theta}(h)$ around $R(h,c)$ per $h$ via Hoeffding's inequality, then applying a union bound over $\mathcal{H}$. Thus, we can have the theorem as
\begin{theorem}[Noiseless finite-class sample complexity]
If the following is satisfied
\begin{eqnarray}
m \ge \frac{1}{\varepsilon} \ln \frac{\abs{\mathcal{H}}}{\delta},
\label{eq:mb-noiseless}
\end{eqnarray}
then with probability at least $1-\delta$ the empirical risk minimizer over $\mathcal{H}$ attains $R(h,c) \le \varepsilon$.
\label{thm:noiseless}
\end{theorem}

In the random classification noise (RCN) model~\cite{AngluinLaird1988}, we consider that each label is flipped independently with probability $\eta \in [0, \frac{1}{2})$:
\begin{eqnarray}
y_j = c(x_j) \oplus \mathscr{B}(\eta),
\label{eq:rcn-model}
\end{eqnarray}
where $\mathscr{B}(\eta) \in \{0, 1\}$ is a Bernoulli random binary number with the error probability $\eta$. Equivalently, the observed label process is a binary symmetric channel with crossover probability $\eta$ acting on the clean labels. The RCN model induces a simple affine relation between the noisy and true risks, which explains the $(1-2\eta)^2$ degradation. Specifically, under Eq.~(\ref{eq:rcn-model}), we have
\begin{eqnarray}
\Pr\bigl( h(x) \neq y \bigr) = \eta + (1-2\eta) \Pr \bigl( h(x) \neq c(x) \bigr).
\end{eqnarray}
This identity reduces the analysis to uniform concentration of bounded i.i.d. averages around their means, but with an \emph{effective margin contraction} by $(1-2\eta)$. We then have the following theorem~\cite{AngluinLaird1988}:
\begin{theorem}[RCN sample complexity]\label{thm:noisy}
If the following is satisfied
\begin{eqnarray}
m \ge \frac{2}{\varepsilon^{2}(1-2\eta)^{2}} \ln \frac{2\abs{\mathcal{H}}}{\delta},
\label{eq:mb-noisy}
\end{eqnarray}
then with probability at least $1-\delta$ the empirical risk minimizer over $\mathcal{H}$ satisfies $R(h,c) \le \varepsilon$.
\end{theorem}
\noindent While the constants above are conservative (Hoeffding-union bounds are rarely tight), their scaling is sharp in the sense that the label noise necessarily induces a $1/\varepsilon^{2}$ dependence and degrades the effective signal-to-noise ratio by $(1-2\eta)$. 

Now, it is convenient to invert Eq.~(\ref{eq:mb-noisy}) to state an explicit confidence guarantee for fixed $m$~\cite{Bang2025}:
\begin{corollary}[Lower bound on achievable confidence]\label{cor:delta-lower}
For any $m, \varepsilon, \eta$ and finite $\mathcal{H}$,
\begin{eqnarray}
\Pr \bigl( R(h,c) \le \varepsilon \bigr) \ge  1 - 2\abs{\mathcal{H}} e^{-\frac{1}{2} \varepsilon^{2} (1-2\eta)^{2} m}.
\label{eq:success-lb-explicit}
\end{eqnarray}
Equivalently, the analysis certifies the following upper bound on the failure probability:
\begin{eqnarray}
\delta_{\min}(m,\varepsilon,\eta,\abs{\mathcal{H}}) = 2\abs{\mathcal{H}} e^{-\frac{1}{2} \varepsilon^{2} (1-2\eta)^{2} m}.
\label{eq:delta-min}
\end{eqnarray}
Suppressing the complexity factor to emphasize the dependence on $(m, \varepsilon, \eta)$, we may summarize the scaling as
\begin{eqnarray}
\Pr \bigl( R(h,c) \le \varepsilon \bigr) \gtrsim 1 - e^{-\gamma m}, \quad \gamma = \frac{1}{2} \varepsilon^{2} (1-2\eta)^{2},
\label{eq:gamma-scaling}
\end{eqnarray}
where $\gtrsim$ hides multiplicative constants in the failure term.
\end{corollary}

In our study, Eq.~(\ref{eq:success-lb-explicit}) and Eq.~(\ref{eq:delta-min}) are used as a \emph{design constraint}: for a fixed $m$, a target failure probability $\delta^\star$ smaller than $\delta_{\min}(m,\varepsilon,\eta,\abs{\mathcal{H}})$ cannot be certified by this concentration analysis without increasing $m$ or reducing $\varepsilon$ or $\eta$.

\section{Learning probability and a finite-budget success criterion}\label{Sec:LearningP}

We now adopt and extend the learning-probability viewpoint of Ref.~\cite{Bang2025} to include an explicit finite-budget stopping event. A trial consumes one new labeled sample and ``time'' is the cumulative number of samples consumed by the protocol.

Fix $\varepsilon\in(0,\frac{1}{2})$ and a sequential learner $L$ with a pre-specified halting rule. Its stopping time is
\begin{eqnarray}
T_L:=\inf\bigl\{t\ge 1:\text{the halting rule of $L$ fires after sample $t$}\bigr\}.
\label{eq:generic-stopping}
\end{eqnarray}
Let $h_{T_L}$ denote the hypothesis returned at $T_L$. The finite-budget learning-success event is
\begin{eqnarray}
\mathcal{E}_{m, \varepsilon} := \bigl\{ T_L\le m~\text{and}~R(h_{T_L},c)\le\varepsilon \bigr\}.
\label{eq:learning-event}
\end{eqnarray}
\begin{definition}[Learning probability]\label{def:PL}
The learning probability of $L$ at sample budget $m$ and tolerance $\varepsilon$ is
\begin{eqnarray}
P_L(m, \varepsilon) := \Pr\bigl( \mathcal{E}_{m, \varepsilon} \bigr),
\label{eq:PL-def}
\end{eqnarray}
where the probability is over the sample sequence and any internal randomness of $L$.
\end{definition}
\noindent Because the events $\mathcal{E}_{m,\varepsilon}$ are nested as either $m$ or $\varepsilon$ increases, $P_L(m,\varepsilon)$ is nondecreasing in both arguments.

The learning probability directly encodes a finite-budget PAC requirement:
\begin{proposition}[PAC-within-budget formulation]
\label{prop:PAC-equiv}
For any sequential learner $L$, sample budget $m$, and inaccuracy $\varepsilon$, the following are equivalent:
\begin{enumerate}
\item $L$ halts by $m$ and returns an $\varepsilon$-approximate hypothesis with probability at least $1-\delta$.
\item $P_L(m,\varepsilon) \ge 1-\delta$.
\end{enumerate}
If $L$ always halts and returns a hypothesis by sample $m$, this reduces to the ordinary fixed-sample $(\varepsilon,\delta)$--PAC statement.
\end{proposition}

Before the halting event is imposed, the concentration analysis yields a conservative but explicit certificate on the training-success probability $P_{\mathrm{train}}(m,\varepsilon):=\Pr[R(h_m,c)\le\varepsilon]$. For finite $\mathcal{H}$ and noise rate $\eta\in[0,\frac{1}{2})$, Eqs.~(\ref{eq:success-lb-explicit}) and (\ref{eq:delta-min}) give
\begin{eqnarray}
P_{\mathrm{train}}(m,\varepsilon) \ge 1 - 2\abs{\mathcal{H}} e^{-\frac{1}{2} \varepsilon^2 (1-2\eta)^2 m} \equiv 1-\delta_{\min}(m,\varepsilon,\eta,\abs{\mathcal{H}}).
\label{eq:PL-erm}
\end{eqnarray}
When capacity terms are immaterial to the comparison, we will also use the simplified \emph{exponential-rate} surrogate
\begin{eqnarray}
P_{\mathrm{BL}}(m,\varepsilon,\eta) := 1 - e^{-\gamma(\varepsilon,\eta) m},
\label{eq:PL-exp}
\end{eqnarray}
where $\gamma(\varepsilon,\eta)=\frac{1}{2}\varepsilon^2(1-2\eta)^2$, which matches the scaling in Eq.~(\ref{eq:gamma-scaling}).

In our secure-learning scenario, $(\varepsilon^\star,\delta^\star)$ is the finite-budget statistical target: $\varepsilon^\star$ is the maximum tolerable population risk and $\delta^\star$ is the maximum failure probability of attaining that risk and halting within $m$. Let $P_{L,\Phi}$ denote the learning probability when data are transmitted through a path $\Phi$.
\begin{definition}\label{def:security-check}
At target $(\varepsilon^\star,\delta^\star)$ and budget $m$, the pair $(L,\Phi)$ satisfies the statistical acceptance condition iff
\begin{eqnarray}
P_{L,\Phi}(m, \varepsilon^\star) \ge 1-\delta^\star.
\end{eqnarray}
\end{definition}
Thus, $P_L(m,\varepsilon)$ is an operational handle for a joint accuracy-and-completion target. Its monotonicity makes the feasible region upward closed in $m$ and $\varepsilon$. The training component remains noise-aware because $\gamma(\varepsilon,\eta)\downarrow0$ as $\eta\uparrow\frac12$. When $P_L$ is estimated from $N$ independent protocol runs, binomial confidence intervals, such as Clopper--Pearson intervals~\cite{Clopper1934}, can turn the empirical success fraction into a lower confidence bound.

\section{Noise--aware certification, halting design, and PAC-security integration}\label{Sec:securePAC}

To incorporate a finite completion requirement into the PAC scenario, we adopt a run-based gate that halts once $M_H$ consecutive validation successes have been observed. The training and validation sets are independent. After the training phase, the hypothesis $h$ is frozen throughout validation, so the validation outcomes are i.i.d. conditional on $h$. The run event is used to control completion within a prescribed budget; it is not, by itself, asserted to be an anytime-valid test of the clean population risk. Statistical reliability upon halting will instead follow from the joint PAC-within-budget event derived in Sec.~\ref{Sec:securePAC}.

The learner fixes $(\varepsilon^\star,\delta^\star,M_H)$ before interacting with a data path $\Phi$. We also introduce the classical design parameter $\bar\eta\in[0,\frac12)$, the largest RCN corruption rate for which the path is regarded as admissible. Given this design, we ask whether the trained hypothesis is $\varepsilon^\star$-approximate and whether the validation gate halts within a finite total budget $m$ whenever $\eta(\Phi)\le\bar\eta$. The symbol $\bar\eta$ is deliberately distinguished from the protocol-derived quantum threshold introduced only in Sec.~\ref{Sec:Quantum_securePAC}.

\subsection{Pre-set halting condition} 

Each validation trial $t$ produces a binary outcome $s_t\in\{0,1\}$ indicating whether the frozen hypothesis passes a fresh label check, with ``$0:=\mathrm{fail}$'' and ``$1:=\mathrm{success}$''~\cite{Lee2021}. The validation stopping time is
\begin{eqnarray}
T_{M_H}:=\inf\bigl\{t\ge M_H:s_{t-M_H+1}=\cdots=s_t=1\bigr\}.
\label{eq:run-stopping-time}
\end{eqnarray}

Let $\epsilon:=R(h,c)\in[0,\frac12]$ denote the clean population risk of the frozen hypothesis. Under RCN with flip rate $\eta\in[0,\frac12)$, a single validation passes with probability
\begin{eqnarray}
q_{\mathrm{obs}}(\epsilon, \eta) := \Pr(s=1) = 1 - \eta - (1 - 2\eta)\epsilon,
\label{eq:qobs-eta}
\end{eqnarray}
which is monotone decreasing in $\epsilon$ for fixed $\eta$, since $\partial q_{\mathrm{obs}}/\partial\epsilon=-(1-2\eta)\le0$. For a fixed window of length $M_H$, the probability that all outcomes pass is $q_{\mathrm{obs}}(\epsilon,\eta)^{M_H}$. For the sequential event $T_{M_H}\le m_{\mathrm{cert}}$, however, repeated opportunities for a run must be included; its probability is therefore the finite-horizon run probability analyzed below and in \ref{appendix:exact-halting}, rather than the single-window quantity $q^{M_H}$.

\subsection{Halting feasibility under a finite budget} 

We now quantify whether the pre-set run gate halts within a finite validation budget.
\begin{lemma}[Run-length mean]\label{lem:run-mean}
Let $q:=q_{\mathrm{obs}}(\epsilon,\eta)$ denote the per-trial success probability of the frozen hypothesis during validation. Then, for any $q\in(0,1)$ and integer $M_H \ge 1$, the expected number of trials to obtain a run of $M_H$ consecutive successes is
\begin{eqnarray}
\mathbb{E}[T_{M_H}] = \frac{1 - q^{M_H}}{(1-q) q^{M_H}},
\label{eq:runlength-mean}
\end{eqnarray}
which scales as $\mathbb{E}[T_{M_H}]\sim q^{-M_H}$ when $q < 1$. Here $\mathbb{E}[T_{M_H}]$ is strictly increasing in $M_H$ and strictly decreasing in $q$, so higher noise or a smaller accuracy target that reduces $q$ increases the validation effort exponentially in $M_H$.
\end{lemma}

\begin{proof}---Introduce the state variable $k \in \{0,1,\dots,M_H\}$ for the current streak length of consecutive successes, and let $E_k$ be the expected additional number of trials needed to reach $M_H$ starting from $k$. Clearly, $E_{M_H}=0$. For $0 \le k < M_H$, after one new trial there is a success with probability $q$ which increases the streak to $k+1$, and a failure with probability $1-q$ which resets the streak to $0$. Hence, we can write
\begin{eqnarray}
E_k = 1 + q E_{k+1} + (1-q) E_0.
\end{eqnarray}
To solve this system explicitly, subtract $E_0$ from both sides and define $F_k:=E_k-E_0$. Then, for $0\le k<M_H$, we have
\begin{eqnarray}
F_k = 1 + q F_{k+1},
\end{eqnarray}
and $F_{M_H} = -E_0$. Unrolling the recursion, we find $F_k = \sum_{j=0}^{M_H-k-1} q^j + q^{M_H-k}\,F_{M_H}$. Setting $k=0$ yields 
\begin{eqnarray}
F_0=\sum_{i=0}^{M_H-1} q^i - q^{M_H} E_0 = 0.
\end{eqnarray}
which directly gives Eq.~(\ref{eq:runlength-mean}).
\end{proof}

For design purposes, many applications require an explicit lower bound on the probability of halting within $m_{\mathrm{cert}}$ trials. Thus, we provide a conservative closed-form lower bound of the halting probability. To this end, partition $m_{\mathrm{cert}}$ into $\lfloor m_{\mathrm{cert}}/M_H\rfloor$ disjoint blocks of length $M_H$. Here, the success in any block means every trial in the block succeeds. Since disjoint blocks comprise independent sets of trials, we obtain the following proposition:
\begin{proposition}[Block lower bound for halting]
\label{prop:block-lb}
With per-trial success probability $q$, the probability of obtaining at least one run of $M_H$ consecutive successes within $m_{\mathrm{cert}}$ trials satisfies
\begin{eqnarray}
\Pr\bigl[T_{M_H}\le m_{\mathrm{cert}}\bigr] \ge 1 - \Bigl(1 - q^{M_H}\Bigr)^{\lfloor m_{\mathrm{cert}}/M_H \rfloor}.
\label{eq:block-lb}
\end{eqnarray}
\end{proposition}

\begin{proof}---In each disjoint block, the event all $M_H$ successes occurs with probability $q^{M_H}$. The independence across the blocks implies that the probability of no all-success block is $\bigl(1-q^{M_H}\bigr)^{\lfloor m_{\mathrm{cert}}/M_H\rfloor}$. Taking the complement yields Eq.~(\ref{eq:block-lb}). This bound considers only nonoverlapping windows, hence it is conservative. It is monotone in $m_{\mathrm{cert}}$ and approaches $1$ as $m_{\mathrm{cert}} \to \infty$ for any fixed $q > 0$.
\end{proof}

We note that the exact halting probability can be computed via dynamic recursion (see~\ref{appendix:exact-halting}).

\subsection{Integrating training and validation}  

Following the previous analysis, we split the overall budget as
\begin{eqnarray}
m = m_{\mathrm{train}} + m_{\mathrm{cert}}.
\label{eq:m_split}
\end{eqnarray}
Let
\begin{eqnarray}
G:=\bigl\{R(h,c)\le\varepsilon^\star\bigr\}, \quad H:=\bigl\{T_{M_H}\le m_{\mathrm{cert}}\bigr\},
\label{eq:G-H-events}
\end{eqnarray}
where $h$ is trained using the first $m_{\mathrm{train}}$ samples and is then frozen. We define the two-phase learning probability by
\begin{eqnarray}
P_{L,\Phi}\bigl(m;\varepsilon^\star,M_H\bigr):=\Pr(G\cap H).
\label{eq:joint-learning-probability}
\end{eqnarray}
The training phase aims to produce an $\varepsilon^\star$-approximate hypothesis with probability at least $1-\delta_{\mathrm{train}}$. Under the RCN model and a finite concept class, we may certify
\begin{eqnarray}
\fl \Pr(G^c)\le\delta_{\mathrm{train}} := \delta_{\min}\bigl(m_{\mathrm{train}}, \varepsilon^\star, \eta, \abs{\mathcal{H}}\bigr) = 2\abs{\mathcal{H}} \exp\Bigl(-\frac{1}{2} \varepsilon^{\star 2} (1-2\eta)^2 m_{\mathrm{train}}\Bigr)
\label{eq:delta-train}
\end{eqnarray}
by Eq.~(\ref{eq:delta-min}). Conditional on $G$, the independent validation trials have per-trial success probability $q\ge q_{\mathrm{obs}}(\varepsilon^\star,\eta)$ by the monotonicity of Eq.~(\ref{eq:qobs-eta}). Therefore,
\begin{eqnarray}
P_{L,\Phi}\bigl(m;\varepsilon^\star,M_H\bigr) \ge 1-\delta_{\mathrm{train}}-\Pr\bigl(T_{M_H}>m_{\mathrm{cert}}\mid G\bigr),
\label{eq:two-phase-union}
\end{eqnarray}
and {\bf Proposition~\ref{prop:block-lb}} gives the conservative closed form
\begin{eqnarray}
P_{L,\Phi}\bigl(m;\varepsilon^\star,M_H\bigr) \;\ge\; 1 - \delta_{\mathrm{train}} - \Bigl(1 - \bigl[q_{\mathrm{obs}}(\varepsilon^\star,\eta(\Phi))\bigr]^{M_H}\Bigr)^{\lfloor m_{\mathrm{cert}}/M_H\rfloor}.
\label{eq:two-phase-lb}
\end{eqnarray}
The dynamic recursion in \ref{appendix:exact-halting} can replace the block bound to sharpen the completion probability without changing the design flow. Importantly, if the right-hand side of Eq.~(\ref{eq:two-phase-lb}) is at least $1-\delta^\star$, then both
\begin{eqnarray}
\Pr(H)\ge1-\delta^\star, \quad \Pr(G\mid H)\ge1-\delta^\star.
\label{eq:halt-conditional-reliability}
\end{eqnarray}
Thus, the joint event supplies a finite-completion guarantee and an accuracy guarantee conditional on halting; the run alone is not used as an anytime-valid $p$-value. We now instantiate the statistical criterion for the halting-based learner. The pair $(L,\Phi)$ satisfies secure PAC compliance at $(\varepsilon^\star,\delta^\star)$ and budget $m$ if
\begin{eqnarray}
\left\{
\begin{array}{ll}
\eta(\Phi) \le \bar\eta & \text{(classical channel admissibility)}, \\[2pt]
P_{L,\Phi}\bigl(m;\varepsilon^\star,M_H\bigr) \ge 1-\delta^\star & \text{(joint accuracy and completion)}.
\end{array}
\right.
\label{eq:secure-final}
\end{eqnarray}

\subsection{Sufficient budget requirement and a secure PAC decision rule}  

We next derive an explicit sufficient total budget. Define the worst-case admissible validation success probability at the target by
\begin{eqnarray}
q_0 = q_{\mathrm{obs}}(\varepsilon^\star,\bar\eta) = 1-\bar\eta-(1-2\bar\eta)\varepsilon^\star,
\label{eq:q0-def}
\end{eqnarray}
and the block-decay rate
\begin{eqnarray}
s_0 = -\ln\bigl(1-q_0^{M_H}\bigr).
\label{eq:s0-def}
\end{eqnarray}

Allocate the total failure probability via $\alpha\in(0,1)$ as
\begin{eqnarray}
\delta_{\mathrm{train}} \le \alpha\delta^\star, \quad \Pr\bigl(T_{M_H} > m_{\mathrm{cert}}\mid G\bigr) \le (1-\alpha)\delta^\star.
\label{eq:alpha-split}
\end{eqnarray}
Under the finite-class empirical risk minimization certificate in Eq.~(\ref{eq:delta-train}), a sufficient training budget is obtained by solving $\delta_{\mathrm{train}} \le \alpha\delta^\star$ for $m_{\mathrm{train}}$, namely,
\begin{eqnarray}
m_{\mathrm{train}}^{\mathrm{lb}}(\alpha) = \frac{2}{\varepsilon^{\star 2}(1-2\bar\eta)^2} \ln\frac{2\abs{\mathcal{H}}}{\alpha\delta^\star}.
\label{eq:mtrain-lb}
\end{eqnarray}
When capacity terms are immaterial, the exponential-rate surrogate in Eq.~(\ref{eq:PL-exp}) gives
\begin{eqnarray}
m_{\mathrm{train}}^{\mathrm{lb\,(BL)}}(\alpha) = \frac{1}{\gamma^\star} \ln\frac{1}{\alpha\delta^\star}, \quad \gamma^\star = \frac{1}{2}\varepsilon^{\star 2}(1-2\bar\eta)^2,
\label{eq:mtrain-lb-bl}
\end{eqnarray}
which is consistent with Eq.~(\ref{eq:gamma-scaling}). For validation completion, {\bf Proposition~\ref{prop:block-lb}} implies that it suffices to choose $n_{\mathrm{cert}}$ blocks so that the conditional block-failure probability is at most $(1-\alpha)\delta^\star$. Thus, we have
\begin{eqnarray}
n_{\mathrm{cert}}^{\mathrm{lb}}(\alpha) = \left\lceil \frac{1}{s_0}\ln{\frac{1}{(1-\alpha)\delta^\star}} \right\rceil, \quad m_{\mathrm{cert}}^{\mathrm{lb}}(\alpha) = n_{\mathrm{cert}}^{\mathrm{lb}}(\alpha) \times M_H,
\label{eq:mcert-lb}
\end{eqnarray}
since $\bigl(1-q_0^{M_H}\bigr)^{n_{\mathrm{cert}}}\le (1-\alpha)\delta^\star$ is equivalent to $n_{\mathrm{cert}}\ge \frac{1}{s_0}\ln{\frac{1}{(1-\alpha)\delta^\star}}$~\footnote{Using the exact block-decay $s_0 = -\ln(1-q_0^{M_H})$ improves over the inequality $-\ln(1-x)\ge x$. The latter yields a simpler but looser bound with $m_{\mathrm{cert}}^{\mathrm{lb}} = M_H \lceil q_0^{-M_H} \ln\frac{1}{(1-\alpha)\delta^\star}\rceil$.}.

Now a sufficient end-to-end budget is given by
\begin{eqnarray}
m_{\mathrm{lb}}(\alpha) = m_{\mathrm{train}}^{\mathrm{lb}}(\alpha) + m_{\mathrm{cert}}^{\mathrm{lb}}(\alpha)
\label{eq:m-lb-alpha}
\end{eqnarray}
To optimize the split $\alpha$, we temporarily ignore the ceiling and set
\begin{eqnarray}
\mathcal{A} = \frac{2}{\varepsilon^{\star 2}(1-2\bar\eta)^2}, \quad \mathcal{B} = \frac{M_H}{s_0}.
\label{eq:A-B-def}
\end{eqnarray}
Then, we get
\begin{eqnarray}
m_{\mathrm{lb}}(\alpha) = \mathcal{A} \ln\frac{2\abs{\mathcal{H}}}{\alpha\delta^\star} + \mathcal{B} \ln\frac{1}{(1-\alpha)\delta^\star}.
\end{eqnarray}
The derivative with respect to $\alpha$ is 
\begin{eqnarray}
\frac{d}{d\alpha}m_{\mathrm{lb}}(\alpha) = -\frac{\mathcal{A}}{\alpha} + \frac{\mathcal{B}}{1-\alpha}.
\end{eqnarray}
 Setting it to zero yields the unique interior minimizer
\begin{eqnarray}
\alpha^\star = \frac{\mathcal{A}}{\mathcal{A} + \mathcal{B}},
\label{eq:alpha-star}
\end{eqnarray}
and strict convexity follows from $\frac{d^2}{d\alpha^2}m_{\mathrm{lb}}(\alpha) = \frac{\mathcal{A}}{\alpha^2} + \frac{\mathcal{B}}{(1-\alpha)^2} > 0$. Substituting $\alpha^\star$ gives the approximately minimized sufficient budget
\begin{eqnarray}
m_{\mathrm{lb}}^{\mathrm{opt}} = \mathcal{A} \ln\frac{2\abs{\mathcal{H}}}{\delta^\star} + \mathcal{B}\ln\frac{1}{\delta^\star} + \mathcal{A}\ln\Bigl(1+\frac{\mathcal{B}}{\mathcal{A}}\Bigr) + \mathcal{B}\ln\Bigl(1+\frac{\mathcal{A}}{\mathcal{B}}\Bigr).
\label{eq:m-lb-opt}
\end{eqnarray}
Here, $m_{\mathrm{lb}}^{\mathrm{opt (BL)}}$ using the exponential-rate surrogate is obtained by replacing the first term $\mathcal{A}\ln\frac{2\abs{\mathcal{H}}}{\delta^\star}$ by $\gamma^{\star-1}\ln\frac{1}{\delta^\star}$ while keeping the same $\alpha^\star$ because the $\alpha$-dependence is governed by the coefficients of $\ln(1/\alpha)$ and $\ln(1/(1-\alpha))$.

Consequently, we obtain the following theorem:
\begin{theorem}[Sufficient budget and secure PAC decision]
\label{thm:budget-lb}
Fix $(\varepsilon^\star,M_H,\delta^\star,\bar\eta)$ and choose some $\alpha\in(0,1)$. If $m_{\mathrm{train}}\ge m_{\mathrm{train}}^{\mathrm{lb}}(\alpha)$ and $m_{\mathrm{cert}}\ge m_{\mathrm{cert}}^{\mathrm{lb}}(\alpha)$, so that $m\ge m_{\mathrm{lb}}(\alpha)$ with the prescribed split, then under the i.i.d. RCN model, independent training and validation sets, and a frozen validation hypothesis, every data path $\Phi$ with $\eta(\Phi)\le\bar\eta$ satisfies
\begin{eqnarray}
P_{L,\Phi}\bigl(m;\varepsilon^\star,M_H\bigr)=\Pr(G\cap H)\ge1-\delta^\star.
\end{eqnarray}
Consequently, $\Pr(H)\ge1-\delta^\star$ and $\Pr(G\mid H)\ge1-\delta^\star$. If a run using these prescribed component budgets fails to halt, one may reject the composite null consisting of channel admissibility together with the stated sampling and protocol assumptions with Type-I error at most $\delta^\star$. Such a failure does not, by itself, identify excessive channel noise as the cause. Equation~(\ref{eq:m-lb-opt}) is the continuously optimized approximation obtained before integer rounding.
\end{theorem}

\section{Quantum-secure PAC learning}\label{Sec:Quantum_securePAC}

In the classical construction, $\bar\eta$ is a designer-chosen tolerance on stochastic label corruption; by itself it places no physical upper bound on information copied by an eavesdropper. We now specialize the same statistical budget law to a BB84-like quantum label path. The quantum layer is not claimed to improve distribution-free PAC sample complexity~\cite{ArunachalamDeWolf2018}. Its role is different and essential for security: nonorthogonal state encoding and information--disturbance relations allow an observed QBER to constrain an eavesdropper's information under a stated protocol and attack model. We denote the resulting protocol-specific threshold by $\eta_{\mathrm{BB84}}$, keeping it distinct from the classical design tolerance $\bar\eta$.

\subsection{Quantum data path protocol $\Phi_Q$}  

We consider a qubit-label-encoded data path $\Phi_Q$. Each binary label $y_j\in\{0,1\}$ is encoded in one of the two BB84 bases, $B_Z=\{\ket{0},\ket{1}\}$ and $B_X=\{\ket{+},\ket{-}\}$, selected uniformly at random. The learner measures each incoming ideal single-qubit signal in an independently selected basis. Basis choices are then announced over an authenticated classical channel, and only basis-matched outcomes are retained~\cite{Song2021a}. The analysis assumes trusted preparation and measurement devices, no unmodeled side channels, a memoryless and basis-symmetric channel, asymptotically many signals, collective attacks, and ideal one-way reconciliation. It does not cover coherent attacks, finite-key corrections, multi-photon implementations, or device-independent security.

If $m_{\mathrm{raw}}$ denotes the number of channel uses, the expected number of sifted samples is $m\approx\kappa m_{\mathrm{raw}}$, with $\kappa=\frac12$ for uniformly and independently selected bases. Hence $m_{\mathrm{train}}$ and $m_{\mathrm{cert}}$ in Eq.~(\ref{eq:m_split}) are sifted-sample budgets, while $m_{\mathrm{train,raw}}\approx\kappa^{-1}m_{\mathrm{train}}$ and $m_{\mathrm{cert,raw}}\approx\kappa^{-1}m_{\mathrm{cert}}$ are expected raw allocations. The operational noise $\eta(\Phi_Q)$ is the QBER estimated from a randomly disclosed parameter-estimation subset after sifting. Under the assumptions above, this disturbance constrains Eve's information as described next. A QBER at or above the derived threshold means only that the present protocol and proof do not certify a positive information advantage; it does not, by itself, prove a particular amount of actual leakage.

\subsection{Information-theoretic meaning of the security} 

Let $Y$ be the uniformly distributed sifted label bit, $B$ the learner's corresponding classical register after measurement and reconciliation, and $E$ Eve's quantum system. Eve's accessible information is
\begin{eqnarray}
I_{\mathrm{acc}}(Y;E):=\sup_{\{M_z\}} I(Y;Z)\le\chi(Y;E),
\label{eq:accessible-holevo}
\end{eqnarray}
where the supremum is over Eve's measurements $\{M_z\}$ and the inequality is the Holevo bound~\cite{Holevo2011book}. For the asymptotic one-way BB84 security proof against collective attacks, the standard complementarity bound gives~\cite{Scarani2009}
\begin{eqnarray}
I(Y;B)=1-h(\eta), \quad \chi(Y;E)\le h(\eta),
\label{eq:eve-holevo}
\end{eqnarray}
where $h(\eta):=-\eta\log_2\eta-(1-\eta)\log_2(1-\eta)$ is the binary entropy. Consequently,
\begin{eqnarray}
I(Y;B)-\chi(Y;E)\ge D_{\mathrm{BB84}}(\eta):=1-2h(\eta).
\label{eq:D-eta}
\end{eqnarray}
Define $\eta_{\mathrm{BB84}}\in(0,\frac12)$ as the unique root of $D_{\mathrm{BB84}}(\eta)=0$. Numerically,
\begin{eqnarray}
\eta_{\mathrm{BB84}} \simeq 0.1100.
\label{eq:bb84-threshold}
\end{eqnarray}
Since $h$ is strictly increasing on $[0,\frac12]$,
\begin{eqnarray}
\eta<\eta_{\mathrm{BB84}}\Longleftrightarrow D_{\mathrm{BB84}}(\eta)>0, \quad \eta\ge\eta_{\mathrm{BB84}}\Longleftrightarrow D_{\mathrm{BB84}}(\eta)\le0.
\label{eq:equivalence}
\end{eqnarray}
Thus, below the threshold the proof certifies the strict information advantage
\begin{eqnarray}
I(Y;B)>\chi(Y;E)\ge I_{\mathrm{acc}}(Y;E).
\label{eq:information-advantage}
\end{eqnarray}
At or above the threshold, $D_{\mathrm{BB84}}(\eta)\le0$ means that this lower bound no longer certifies a positive advantage. It is not a universal impossibility statement about every quantum protocol, learner, or attack realization.

This distinction explains the importance of the quantum layer. For a classical crossover channel, a copied symbol need not disturb the learner's copy, so an observed error rate alone does not generally bound an eavesdropper's information. In the BB84-like path, nonorthogonal encoding makes information acquisition physically linked to disturbance, and the measured QBER therefore supports the protocol-specific bound in Eq.~(\ref{eq:eve-holevo}). The resulting guarantee is an information-advantage admissibility condition complementing the PAC statement; it is not, by itself, composable semantic security. Achieving end-to-end confidentiality would additionally require the usual parameter-estimation, error-correction, privacy-amplification, authentication, and finite-key analysis, which lie outside the present learning model.

\subsection{Quantum-secure PAC condition}  

Combining the protocol-specific information condition with {\bf Theorem~\ref{thm:budget-lb}} yields the following one-sided sufficient result.
\begin{theorem}[Quantum-admissible sufficient budget law; specialization of {\bf Theorem~\ref{thm:budget-lb}}]
\label{thm:quantum-pac}
Adopt the two-phase learner $L$ with pre-set targets $(\varepsilon^\star,\delta^\star,M_H)$ and independent training and validation data. Let $\Phi_Q$ be the BB84-like qubit path under the assumptions stated in Sec.~\ref{Sec:Quantum_securePAC}, and let $\eta(\Phi_Q)$ be its asymptotic sifted QBER. For a conservative statistical design, set $\bar\eta=\eta_{\mathrm{BB84}}$ in Eqs.~(\ref{eq:q0-def}), (\ref{eq:s0-def}), and (\ref{eq:A-B-def}). If
\begin{eqnarray}
\eta(\Phi_Q)<\eta_{\mathrm{BB84}},
\label{eq:quantum-admissibility}
\end{eqnarray}
then, for any $\alpha\in(0,1)$, the sufficient budgets in Eqs.~(\ref{eq:mtrain-lb}) and (\ref{eq:mcert-lb}) give
\begin{eqnarray}
P_{L,\Phi_Q}\bigl(m;\varepsilon^\star,M_H\bigr)=\Pr(G\cap H)\ge1-\delta^\star,
\label{eq:quantum-joint-guarantee}
\end{eqnarray}
and the quantum path simultaneously satisfies $I(Y;B)>\chi(Y;E)\ge I_{\mathrm{acc}}(Y;E)$. Ignoring integer rounding, the confidence split is minimized at $\alpha^\star=\mathcal{A}/(\mathcal{A}+\mathcal{B})$, and the resulting sufficient sifted-sample budget is
\begin{eqnarray}
m \ge m_{\mathrm{lb}}^{\mathrm{opt}} = \mathcal{A}\ln\frac{2\abs{\mathcal{H}}}{\delta^\star} + \mathcal{B}\ln\frac{1}{\delta^\star} + \mathcal{A}\ln\Bigl(1+\frac{\mathcal{B}}{\mathcal{A}}\Bigr) + \mathcal{B}\ln\Bigl(1+\frac{\mathcal{A}}{\mathcal{B}}\Bigr).
\label{eq:mlbopt-q}
\end{eqnarray}
For uniform random bases, provisioning $m_{\mathrm{raw}}\ge2m_{\mathrm{lb}}^{\mathrm{opt}}$ gives the corresponding expected raw channel-use budget.
\end{theorem}

\begin{proof}---For every $\eta(\Phi_Q)<\eta_{\mathrm{BB84}}=\bar\eta$, {\bf Theorem~\ref{thm:budget-lb}} applies with the worst-case statistical design value $\bar\eta$. The training term is bounded by $\alpha\delta^\star$, and the conditional validation noncompletion term is bounded by $(1-\alpha)\delta^\star$, proving Eq.~(\ref{eq:quantum-joint-guarantee}). Equation~(\ref{eq:information-advantage}) independently supplies the quantum information-advantage condition. Minimizing the same sufficient budget over $\alpha$ gives Eq.~(\ref{eq:mlbopt-q}); basis matching gives the expected factor of two in raw transmissions.
\end{proof}

The result deliberately separates two layers. The PAC layer controls the probability of obtaining an accurate hypothesis and completing validation within budget. The quantum layer constrains Eve's accessible information from an observable disturbance under a specified physical model. The Holevo bound does not tighten the PAC confidence or lower the learning sample complexity. Its importance is that a comparable inference about copied information cannot be made from the crossover rate of an unconstrained classical path. If $\eta(\Phi_Q)\ge\eta_{\mathrm{BB84}}$, the present BB84 one-way bound does not certify the quantum layer, even though a statistical learning task may remain feasible.

\section{Summary and Discussion}\label{Sec:6}

We have formulated a finite-budget secure PAC criterion around the joint event that training produces an $\varepsilon^\star$-approximate frozen hypothesis and a run-based validation gate halts within the prescribed budget. Under i.i.d. RCN and a finite hypothesis class, the training concentration bound and a conservative finite-horizon run bound yield the sufficient budget in Eq.~(\ref{eq:m-lb-opt}). This formulation corrects an important distinction: $q^{M_H}$ is the probability of one fixed all-success window, whereas completion of a sequential run by a finite horizon must account for all possible windows. The run gate is therefore used for finite completion, while the PAC accuracy statement is carried by the joint event $G\cap H$.

For a classical path, the procedure is to select $(\varepsilon^\star,\delta^\star,M_H)$ and a designer-chosen admissible corruption rate $\bar\eta$, compute $q_0$ and $s_0$, and allocate independent training and validation budgets. Meeting the sufficient budget gives both $\Pr(H)\ge1-\delta^\star$ and $\Pr(G\mid H)\ge1-\delta^\star$. A failure to halt permits rejection of the composite model assumptions at level $\delta^\star$, but does not isolate channel noise as the cause. The quantities called lower-bound budgets through the notation $m_{\mathrm{lb}}$ are conservative sufficient requirements, not information-theoretic converses.

The quantum specialization retains this statistical budget law but changes the meaning of path admissibility. For the stated idealized BB84-like protocol, the Holevo bound gives $I(Y;B)-\chi(Y;E)\ge1-2h(\eta)$ and the protocol-specific threshold $\eta_{\mathrm{BB84}}\simeq0.1100$. Below it, the learner has a certified information advantage over Eve's accessible information. This is the central role of quantum mechanics in the construction: nonorthogonal encoding links information acquisition to observable disturbance, whereas a classical symbol can generally be copied without such a disturbance signature. The quantum layer neither improves PAC sample complexity nor strengthens $\delta^\star$; it supplies a complementary physical information constraint. At or above the threshold, the present proof ceases to certify that constraint rather than proving universal impossibility.

The scope is correspondingly delimited. The classical result assumes i.i.d. random classification noise, a finite hypothesis class, independent validation data, and a hypothesis frozen during validation. The quantum result additionally assumes ideal single-qubit signals, authenticated classical communication, trusted devices, a memoryless basis-symmetric channel, asymptotic parameter estimation, collective attacks, and ideal one-way reconciliation. The raw budget conversion through $\kappa$ is an expectation, and finite-key fluctuations, coherent attacks, implementation side channels, and composable confidentiality are outside the present result. Natural extensions include anytime-valid validation rules, finite-key bounds, stronger attack models, and privacy-amplified protocols that couple learning accuracy to a full cryptographic security definition.

\section*{Acknowledgement}

JB thanks Marcin Paw\l{}owski, Wooyeong Song, and Gwangil Bae for helpful discussions and comments. This work was supported by the Ministry of Science, ICT and Future Planning (MSIP) through the National Research Foundation of Korea (RS-2024-00432214, RS-2025-03532992, and RS-2025-18362970) and the Institute of Information and Communications Technology Planning and Evaluation grant funded by the Korean government (RS-2019-II190003, ``Research and Development of Core Technologies for Programming, Running, Implementing and Validating of Fault-Tolerant Quantum Computing System''), the Korean ARPA-H Project through the Korea Health Industry Development Institute (KHIDI), funded by the Ministry of Health \& Welfare, Republic of Korea (RS-2025-25456722), and the Ministry of Trade, Industry and Resources (MOTIR), Korea, under the project ``Industrial Technology Infrastructure Program'' (RS-2024-00466693). This work is also supported by the Grant No.~K25L5M2C2 at the Korea Institute of Science and Technology Information (KISTI). We acknowledge the Yonsei University Quantum Computing Project Group for providing support and access to the Quantum System One (Eagle Processor), which is operated at Yonsei University.

\appendix

\section{Exact halting probability via dynamic recursion}\label{appendix:exact-halting}

While the lower bound in {\bf Proposition~\ref{prop:block-lb}} is informative, we can compute this probability exactly by a dynamic recursion over the streak. Let $p_t(k)$ be the probability that after $t$ validation trials, the process is in streak $k \in \{0,\dots,M_H-1\}$ without having yet reached $M_H$, and let $Q_t$ be the cumulative probability that the streak $M_H$ has occurred by time $t$. Initialize $p_0(0)=1$ and $p_0(k)=0$ for $k\ge 1$, $Q_0=0$. Then, for $t\ge 0$, we find
\begin{eqnarray}
p_{t+1}(0) &=& (1-q)\sum_{k=0}^{M_H-1} p_t(k) \nonumber \\
p_{t+1}(k) &=& q p_t(k-1) \quad \text{for $1 \le k \le M_H-1$} \nonumber \\
Q_{t+1} &=& Q_t + q p_t(M_H-1).
\label{eq:dp-recursion}
\end{eqnarray}
Here, note that the exact halting probability within $m_{\mathrm{cert}}$ trials is $Q_{m_{\mathrm{cert}}}$. This dynamic program runs in time $O(M_H\,m_{\mathrm{cert}})$ and yields the tightest computable certificate.

\section{Derivation of the BB84 threshold $\eta_{\mathrm{BB84}}\simeq0.1100$}\label{appendix:bb84-threshold}

For completeness, we recall the asymptotic one-way BB84 argument used in Sec.~\ref{Sec:Quantum_securePAC}. For a uniformly distributed sifted bit $Y$ sent through a basis-symmetric binary channel with QBER $\eta$, the learner's mutual information is
\begin{eqnarray}
I(Y;B)=1-h(\eta).
\end{eqnarray}
The standard complementarity-based security proof against collective attacks gives $H(Y\mid E)\ge1-h(\eta)$~\cite{Scarani2009}. Since $H(Y)=1$ and $\chi(Y;E)=H(Y)-H(Y\mid E)$ for the associated classical--quantum state,
\begin{eqnarray}
\chi(Y;E)\le h(\eta).
\end{eqnarray}
Combining the two relations yields
\begin{eqnarray}
I(Y;B)-\chi(Y;E)\ge D_{\mathrm{BB84}}(\eta):=1-2h(\eta).
\end{eqnarray}

On $0<\eta<\frac12$, the binary entropy satisfies
\begin{eqnarray}
h'(\eta)=\log_2\frac{1-\eta}{\eta}>0,
\end{eqnarray}
so $D'_{\mathrm{BB84}}(\eta)=-2h'(\eta)<0$. Moreover, $D_{\mathrm{BB84}}(0)=1$ and $D_{\mathrm{BB84}}(\frac12)=-1$. Continuity and strict monotonicity therefore give a unique zero in $(0,\frac12)$. Solving $h(\eta)=\frac12$ gives
\begin{eqnarray}
\eta_{\mathrm{BB84}}=0.110027\ldots\simeq0.1100.
\end{eqnarray}
It follows that $D_{\mathrm{BB84}}(\eta)>0$ exactly for $\eta<\eta_{\mathrm{BB84}}$. For these QBER values the lower bound certifies $I(Y;B)>\chi(Y;E)\ge I_{\mathrm{acc}}(Y;E)$. When $\eta\ge\eta_{\mathrm{BB84}}$, the lower bound is nonpositive and hence inconclusive about a strict information advantage; no converse or protocol-independent impossibility is asserted.

\section*{References}

\bibliographystyle{iop}

\end{document}